\newcommand{\be}{\begin{equation}}
\newcommand{\ee}{\end{equation}}
\newcommand{\ba}{\begin{eqnarray}}
\newcommand{\ea}{\end{eqnarray}}
\newcommand{\nn}{\nonumber}

\documentclass{ws-procs9x6}

\begin{document}

\title{QCD Phase Diagram at small Baryon Densities from Imaginary $\mu$: Status Report}

\author{O.~PHILIPSEN\footnote{\uppercase{T}alk given by \uppercase{O}.\uppercase{P}.,
\uppercase{PPARC} \uppercase{A}dvanced \uppercase{F}ellow}}

\address{Dept.~of Physics and Astronomy,
University of Sussex, \\
Brighton BN1 9QH, UK\\
E-mail: o.philipsen@sussex.ac.uk}  

\author{PH.~DE FORCRAND}

\address{ETH, CH-8093 Z\"urich, Switzerland, and \\
CERN, Physics Dept.~, TH Unit,
CH-1211 Geneva 23, Switzerland\\ 
E-mail: forcrand@phys.ethz.ch}

\maketitle

\abstracts{
We summarize our recent results for the ($T,\mu$)-phase diagram of $N_f=3$ QCD.
Finding a strong variation of the critical endpoint $\mu_c$ with the quark mass $m$, 
we point out that an endpoint within reach of experiment requires fine-tuned quark masses.
We further discuss
the strategy and first data towards extending our results to the physical $N_f=2+1$ theory.
}

\section{Introduction}

Currently, several approaches are used to side-step the sign problem of lattice QCD in
attempts to compute the low density QCD phase diagram. These are based on multi-parameter
reweighting \cite{fk}, Taylor expansions \cite{allt} and analytic continuation from simulations at imaginary
$\mu$ \cite{im}, for which there is no sign problem \cite{lp}.  
%The last of these is numerically the simplest.
While reweighting is exponentially difficult on large volumes, calculation of Taylor coefficients
becomes increasingly compli\-cated with growing order. By contrast, a simulation at imaginary $\mu=i\mu_I$
is no more difficult than one at $\mu=0$. Moreover, it is the only approach for which two simulation
parameters are varied, resulting in uncorrelated statistical errors for different values of $\mu_I$.
Monte Carlo results for imaginary $\mu$ thus involve no approximations.
Within the circle of analyticity around $\mu=0$, they can be fitted by truncated Taylor series in 
$(\mu_I/T)^2$
and analytically continued to real $\mu$. While truncating the series introduces an approximation,
it is well controlled because it can be compared to the full result at imaginary $\mu$.

The task for lattice simulations is to first map out the phase diagram in the $(m_{u,d},m_s)$ quark mass plane at $\mu=0$, i.e.~to find the critical line $m_s^c(m_{u,d})$ of second order transitions,
separating the first order region from the crossover region. 
In a second step, we wish to know the surface emanating from this line in the $\mu\neq 0$ direction \cite{cs}. So far, even at $\mu=0$ we know only one point of the critical line, corresponding
to degenerate quark masses, $m_c(\mu=0),N_f=3$, and only for one coarse lattice spacing,  $N_t=4$.
In a recent paper we have computed the $\mu\neq 0$ phase diagram for the $N_f=3$ theory and determined the quark mass dependence of the critical endpoint \cite{fp2}. Here we summarize these results and
extend them to the physical case of $N_f=2+1$. The general properties of QCD
with complex chemical potential as well as analyticity properties of observables are discussed in the literature \cite{im,fp2}.
\begin{figure}[t] \label{bc}
\leavevmode
\includegraphics[width=5.3cm]{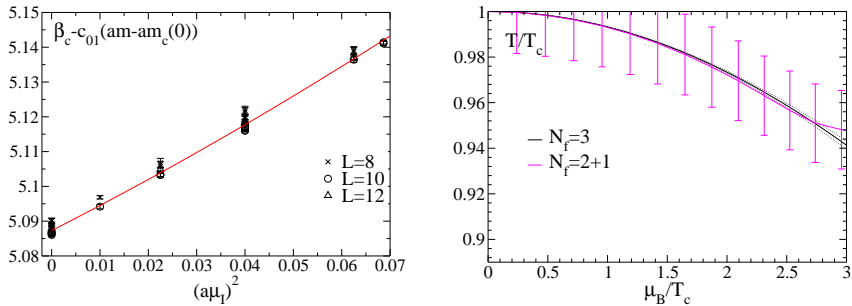}\hspace*{0.5cm}
\includegraphics[width=5.3cm]{philipsen_fig2.eps}
%~\vspace*{-0.3cm}
\caption[]{Left: The pseudo-critical gauge coumpling for various quark masses and lattice sizes 
as a function of imaginary $\mu=i\mu_I$. Right: $N_f=3$ \cite{fp2} and 
$N_f=2+1$ \cite{fk} results.}
\end{figure}

\section{The pseudo-critical line for $N_f=3$}
 
For $N_f=3$, we computed the pseudo-critical gauge couplings at the phase boundary, and their dependence on imaginary chemical potential as well as the quark mass. 
The data are shown in Figure \ref{bc} for different lattice sizes. 
They are well fitted by a polynomial of degree one
in the $(am)$ and two in $(a\mu_I)^2$. We have explicitly checked that additional polynomial
terms are statistically insignificant. 
The figure shows that, to good approximation, the leading $\mu_I^2$-term
suffices over the interval considered, but our accuracy is good enough to also determine the next-to-leading coefficient. Analytic continuation 
%to real $\mu$ 
and conversion to physical units yields
\ba
\frac{T_c(\mu,m)}{T_c(\mu=0,m_c(0))}&=& 1 +
1.94(2)\left( \frac{m-m_c(0)}{\pi T_c(0,m_c)}\right)\nn\\
&&\hspace*{-1cm}+0.602(9)\left(\frac{\mu}{\pi T_c(0,m_c)}\right)^2
+0.23(9)\left(\frac{\mu}{\pi T_c(0,m_c)}\right)^4.
\ea
Note that the natural expansion parameters are $m/(\pi T),\mu^2/(\pi T)^2$, leading to 
Taylor coefficients of order one. This is understood in terms of  
the finite temperature Matsubara modes $\sim \pi T$, which set the energy scale.  
Since the strange quark mass and its difference with $m_c(0)$ are small compared to $\pi T_c$, %$T_c(\mu)$ 
the critical temperature is fairly insensitive to quark masses in the light quark regime. 
It is then sensible to compare our $N_f=3$ results with the $N_f=2+1$ curve of Fodor
and Katz \cite{fk} (Figure \ref{bc} right), and excellent agreement is found. 
Note that $T_c(\mu)$ decreases only at the percent level over the whole range of baryon chemical potentials considered here, which thus has a very weak effect. This is a natural consequence of the
smallness of the effective expansion parameter with Taylor coefficients of order one.
\begin{figure}[t] \label{schem}
\leavevmode
\includegraphics[width=5.3cm]{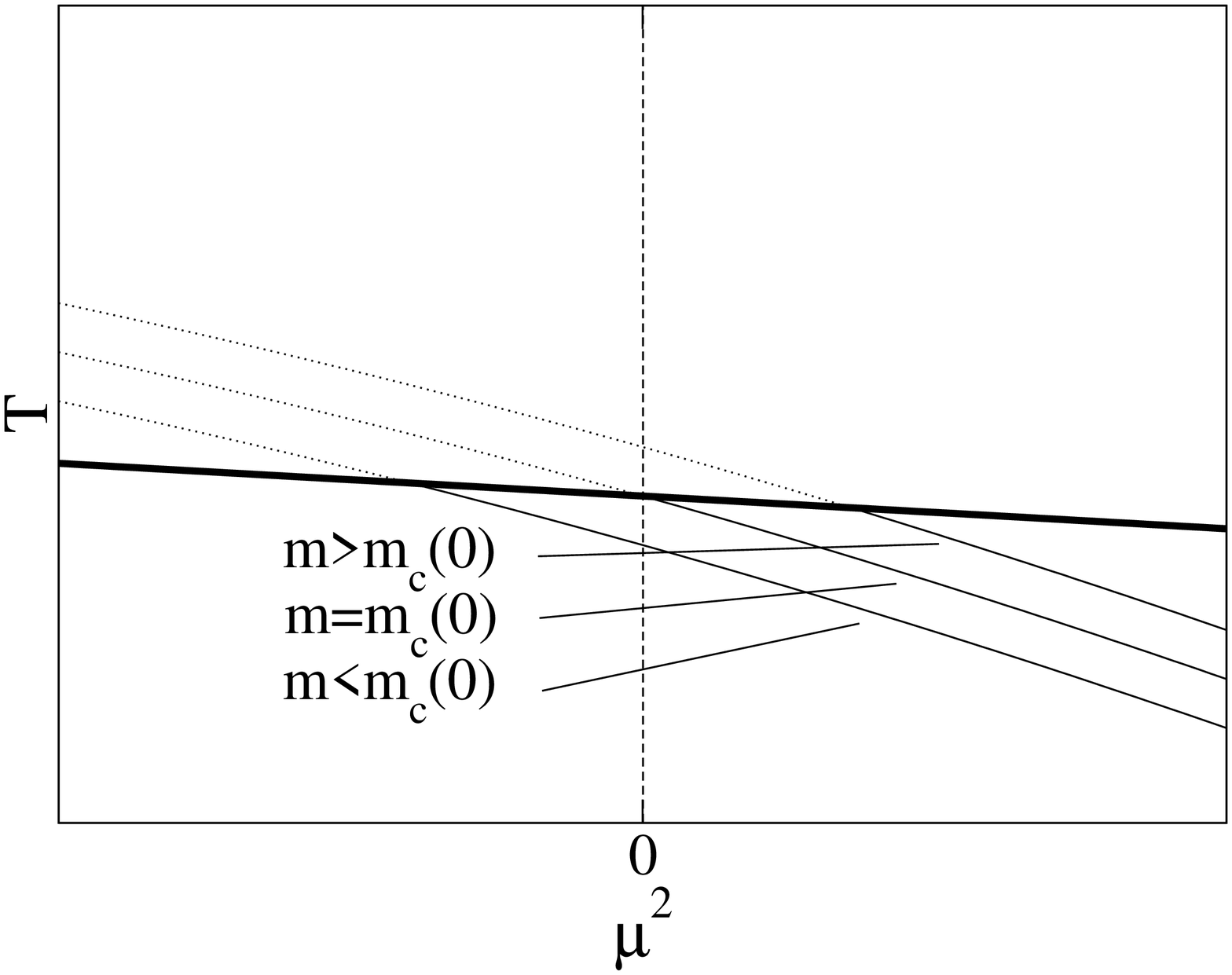}\hspace*{0.5cm}
\includegraphics[width=5.3cm]{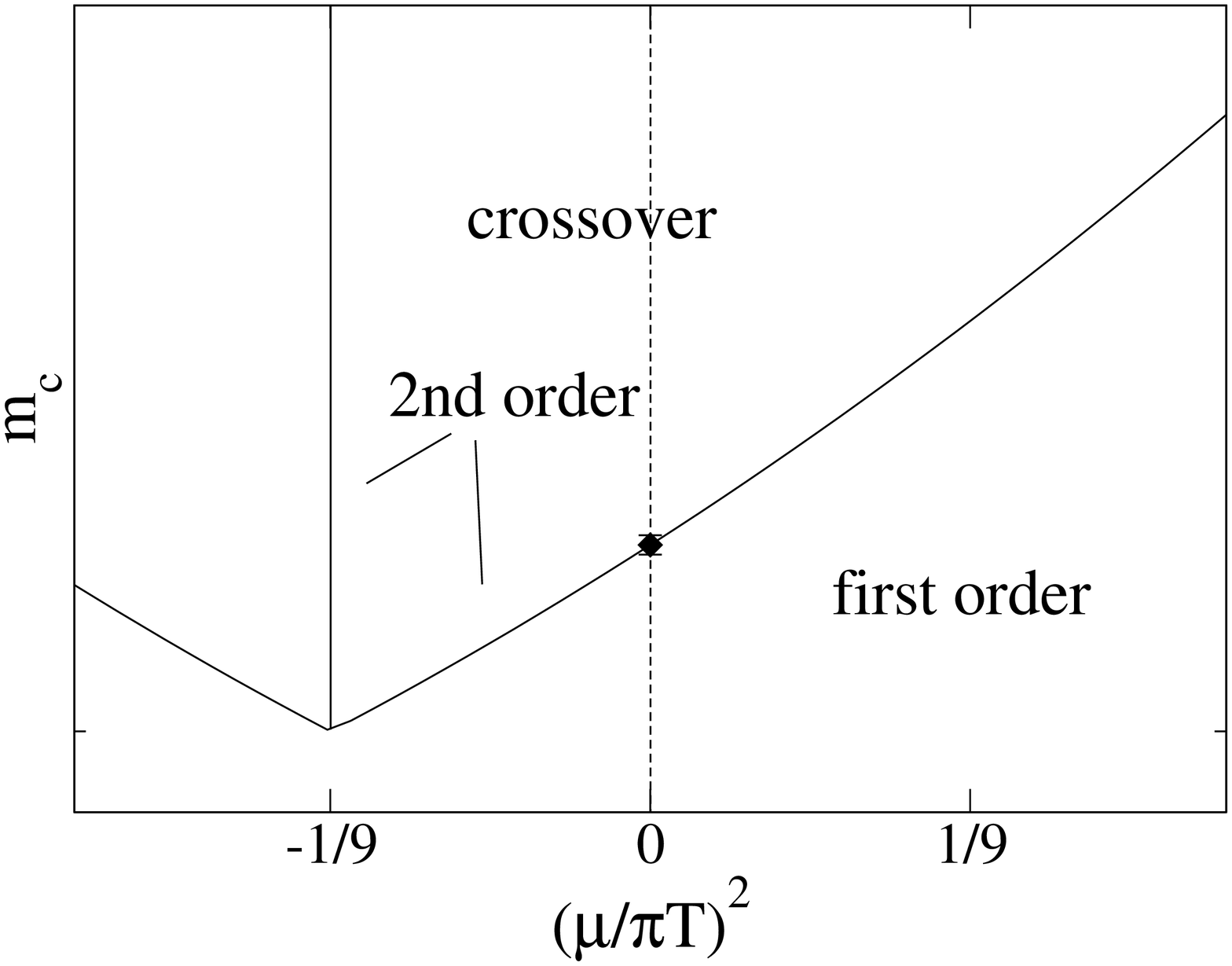}
\caption[]{
Left: Phase diagram for different quark masses,
dotted lines denote crossover.
The bold curve $T^*(\mu^2)$ represents the line of endpoints.
Right:  Lines of critical quark mass separating 
first order and crossover regions.}
\end{figure}

\section{The critical endpoint and its quark mass dependence}

\begin{figure}[t] \label{b4}
\leavevmode
\includegraphics[width=5.3cm]{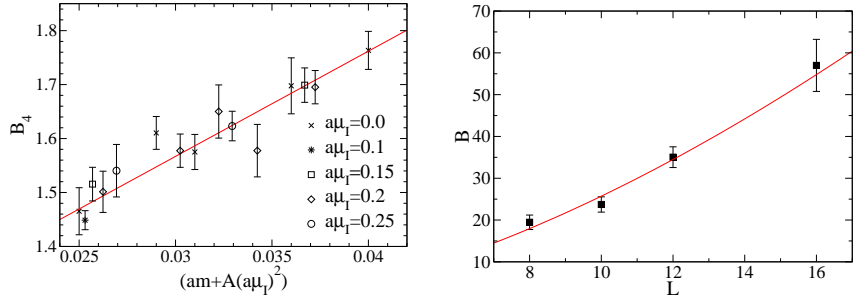}\hspace*{0.5cm}
\includegraphics[width=5.3cm]{philipsen_fig6.eps}
\caption[]{
Left: The Binder cumulant on an L=8 lattice. Right: 
Finite size scaling of the fit parameter $B$. The line corresponds to 3d Ising scaling.}
\end{figure}

In the three flavor theory, we have a 3d parameter space, $(T,\mu,m)$.
The pseudo-critical temperature, $T_c(\mu,m)$, represents a surface in this space.
On this surface there is a line of critical points, $T^*(\mu^2)=T_c(\mu^2,m_c(\mu))$, separating
first order transitions from crossover. Projections of this situation to the $(T,\mu^2),(m,\mu^2)$
planes are  shown in Figure \ref{schem}. As the quark mass is increased from $m_c(0)$, the critical
endpoint moves to a finite $\mu_c$, while a decrease moves it to imaginary $\mu_c$.
The change of $m_c$ with $\mu$, or vice versa, is best seen in the $(m,\mu^2)$-plane,
where it forms a critical line separating regions of first order transitions and crossover.
We determine this line around $\mu=0$ by measuring the Binder cumulant of $\bar{\psi}\psi$.
In the infinite volume limit it assumes a universal value at a 3d Ising critical point,
\be
B_4(m_c,\mu_c)=\frac{\langle(\delta\bar{\psi}\psi)^4\rangle}
{\langle(\delta\bar{\psi}\psi)^2\rangle}\rightarrow 1.604,\quad V\rightarrow \infty,
\ee
while it is smaller or larger in first order and crossover regions, respectively.
Data for various quark masses, as well as its 
universal finite size scaling, are shown in Figure \ref{b4}.
Fitting the data to a Taylor expansion about the critical point, 
$B_4(am,a\mu)=1.604 + B\left(am-am_c(0) + A(a\mu)^2\right) + \ldots$, the parameter
$A$ represents the first coefficient of the series
\be \label{mc}
\frac{m_c(\mu)}{m_c(\mu=0)}=1 + 0.84(36) \left(\frac{\mu}{\pi T}\right)^2+\ldots
\ee
Like for $T_c(\mu)$, the leading Taylor 
coefficient is of order one. This implies that the change in $m_c(\mu)$ is very weak up to chemical
potentials of $\mu_B=3\mu\sim 500$ MeV. Conversely, $\mu_c(m)$
is extremely sensitive to the precise value of the quark mass. We believe that this statement 
holds for all discretizations, because the $\mu$-independent multiplicative mass renormalization drops out of the ratio in (\ref{mc}), which should have only additive $O(a^2)$ corrections.
This means that for growing quark masses $m>m_c(0)$, the critical point $\mu_c(m)$ will rapidly shoot
to very large values. Hence a small $\mu_c$ requires a fine-tuned quark mass.

\section{Towards the $N_f=2+1$ phase diagram}

\begin{figure}[t] \label{m1m2}
\vspace*{-0.3cm}
\centerline{\includegraphics[width=6.0cm]{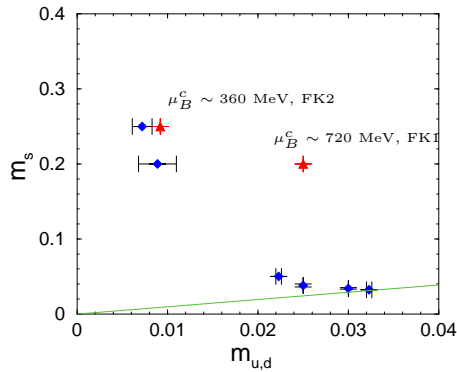}
\put(-110,105){\tiny  $\mu_B^c\sim 360$ MeV, FK2}
\put(-70,90){\tiny  $\mu_B^c\sim 720$ MeV, FK1}
}
\vspace*{-0.3cm}
\caption[]{
Measured critical line separating 1.O. and crossover for $N_f=2+1$ at $\mu=0$ (diamonds),
and finite $\mu$-endpoints of Fodor and Katz \cite{fk} (triangles).}
\end{figure}

In the case of physical interest, we have a 4d parameter space $(T,\mu,m_{u,d},m_s)$.
The critical surface can be determined in a two-step procedure.
First, we need to complete the $(m_{u,d},m_s)$ phase diagram at $\mu=0$, i.e.~find
the line $m_s^c(m_{u,d},\mu=0)$.  The results of this step are shown 
in Figure \ref{m1m2}.
As the light quark mass is decreased, strongly non-linear behavior is encountered, and linear
extrapolations from the $N_f=3$ result to the physical point are clearly ruled out.
In a second step in progress, 
we compute the change of this line with finite $\mu_I$, leading to
the critical surface $m_s^c(m_{u,d},\mu)$.  
It is intriguing to consider the simulation point of Fodor and Katz \cite{fk} 
for physical quark mass values (FK2) in this diagram.
This point is only slightly displaced from
our critical line $m_s^c(m_{u,d},\mu=0)$, but corresponds to $\mu_c\sim 120$ MeV.
This is consistent with our observation of a rapid change of $\mu_c$ for small variations of the 
quark masses. Since quark masses are very cut-off sensitive,
one thus has to expect large shifts of the critical point 
in physical units as the continuum limit is approached.

%%%%%%%%%%%%%%%%%%%%%%%%%%%%%%%%%%%%%%%%%%%%%%%%%%%%%%%%%%%%%%%%%%%%%%%
% 
%Use this if your figures are put in a subdirectory having the same
%name as the main latex file, ie: 
%
%      ws-procs9x6/procs-fig1.eps      
%      ws-procs9x6/procs-fig2.eps      
%      ws-procs9x6/procs-fig3.eps      
%      ws-procs9x6/procs-fig4.eps      
%      etc.
%
%\begin{figure}[htbp] %ORIGINAL SIZE: width=1.4TRUEIN; height=1.5TRUEIN
%\figurebox{}{}{procf1} %100 percent
%\caption{Labeled tree {\it T}.}
%\end{figure}
%
%%%%%%%%%%%%%%%%%%%%%%%%%%%%%%%%%%%%%%%%%%%%%%%%%%%%%%%%%%%%%%%%%%%%%%%

\end{document}